# STAR AND CLUSTER FORMATION IN THE LARGE MAGELLANIC CLOUD

SIDNEY VAN DEN BERGH

Dominion Astrophysical Observatory, Herzberg Institute of Astrophysics,

National Research Council of Canada, 5071 West Saanich Road, Victoria, British Columbia,

V8X 4M6, Canada; vdb@dao.nrc.ca

## ABSTRACT

A great burst of cluster formation increased the rate at which open clusters were formed in the Large Magellanic Cloud 3-5 Gyr ago by at least an order of magnitude. On the other hand the rate of star formation ~ 4 Gyr appears to have increased by a factor of only 2-4. This shows that the rate of cluster formation is not a good tracer of the rate at which stars are formed. Normalized to the same rate of star formation, the Large Cloud is presently forming $\gtrsim$ 600 times more star clusters than the Local Group dwarf irregular IC 1613. The high rate of cluster formation in merging gas-rich galaxies suggests that strong shocks might favor the formation of clusters.

subject headings: galaxies: Magellanic Clouds - galaxies: star clusters - galaxies: starburst

## 1. INTRODUCTION

Baade (1963, p. 230) wrote that "Another remarkable thing about the Large Cloud is the unbelievable number of star clusters that it contains. By contrast, not a single star cluster is found in IC 1613, although there is a large super association." More recently van den Bergh (1979) confirmed Baade's conclusion that IC 1613 is cluster-poor by showing that the Small Magellanic Cloud (SMC), when viewed at the same resolution as IC 1613, contains 15 clusters, whereas none were seen in IC 1613. Recently Hodge (1998) has estimated that, normalized to the same rate of star formation, the Large Magellanic Cloud (LMC) presently forms $>600$ times as many clusters as does IC 1613. This suggests that Baade (1963, p. 231) may have been too optimistic when he wrote "I am sure that the star clusters of the Large Cloud can furnish the history of the whole evolution of at least one stellar system."

## 2. CLUSTER FORMATION HISTORY IN THE LMC

Da Costa (1991) showed that cluster formation in the Large Cloud mainly took place during two great bursts. All of the Large Cloud globular clusters were formed during the first burst, which lasted $\lesssim 1$ Gyr (Johnson et al. 1999) [NGC 1466, NGC 2257, Hodge 11], (Olsen et al. 1999a,b) [NGC 1754, NGC 1835, NGC 1898, NGC 1916, NGC 2005, NGC 2019]. The second great burst of cluster formation started 3-4 Gyr ago, and continues to the present day. Sarajedini (1998, 1999) has recently found 3 LMC star clusters with ages of ~ 5 Gyr. Taken at face value this result suggests that the recent great burst of star formation in the Large Cloud may have "ramped up" for ~ 1 Gyr. Only a single cluster [ESO 121-SC03] is known to have formed during the "dark ages" that lasted from ~ 13 Gyr to ~ 6 Gyr ago. From a survey of old cluster candidates Geisler et al. (1997) conclude that "there are few, if any, genuine old clusters in the LMC left to be found". In their data sample there are 12 clusters with ages in the range 2.0-2.9 Gyr, compared to zero clusters with ages of 4.0-8.9 Gyr. For a cluster formation rate increasing by factors of 10x, 20x and 30x to 12 per Gyr, Poisson statistics yield probabilities of 0.002, 0.05 and 0.14, respectively, for finding zero clusters with ages in the range 4.0 Gyr to 8.9 Gyr. Taken at face value these results suggest that the rate of cluster formation increased by one or two orders of magnitude during the recent great burst of cluster formation in the LMC. This calculation probably over-estimates the size of the increase in the rate of cluster formation because it does not take into account the fact that some clusters which formed ~ 6 Gyr ago may have disintegrated. Such cluster disintegration is thought to be due mainly to interactions with giant molecular clouds (van den Bergh & McClure 1980). However, recent infrared observations by NANTEN (Fukui et al.1999) show that giant molecular clouds are much less frequent in the LMC than they are in the main body of the Galaxy. Therefore cluster destruction may be less important in the LMC than it is in the main body of the Galaxy. Strong observational support for this suggestion comes from the work of Hodge (1988). The fact that the SMC exhibits no evidence for a burst of cluster formation ~ 3 Gyr ago (Da Costa 1991, Mighell, Sarajedini & French 1998, 1999) suggests that the burst of cluster formation in the LMC was probably not triggered by a close tidal interaction between the Large Cloud and the Small Cloud. After excluding clusters with ages < 1 Gyr (for which the data are incomplete) a Kolmogorov-Smirnov test shows that there is only a 5% probability that the cluster age distribution in the LMC (Geisler et al. 1997) and in the SMC (taken from Figure 12 of Mighell, Sarajedini & French 1998) were drawn from the same parent population.

### 3. STAR FORMATION HISTORY OF THE LARGE CLOUD

The hypothesis that no stars (and supernovae!) formed during the "dark ages" between ~ 13 Gyr and ~ 5 Gyr ago seems to be ruled out by the observation that the heavy element abundances in LMC stars and clusters, appears to have increased by ~ 0.7 dex between the beginning and the end of this period. The most conservative interpretation of the observations discussed in § 2 is that the rate of <u>star</u> formation increased less dramatically 3-5 Gyr ago than did the rate of <u>cluster</u> formation. It is noted in passing that a significant number of present-day field stars with ages < 4 Gyr probably formed originally in clusters that have since disintegrated.

The first indication that the rate of star formation did, in fact, increase a few Gyr ago was derived from observations of the color-magnitude diagrams of LMC fields stars by Butcher (1977). He concluded that "the bulk of star formation began 3-5 x $10^9$ years ago in the LMC, rather than 10 x $10^9$ years ago as in the Galaxy". This conclusion was subsequently confirmed by Stryker (1984). For complete references to subsequent work on this subject the reader is referred to Gallagher et al. (1996), Holtzman et al. (1997) and Geha et al. (1998). Geha et al. and Holtzman et al. both conclude that the rate of star formation remained approximately constant during most of the history of the Large Cloud, and that it then increased by a factor of about three some 2 Gyr ago. On their picture roughly half of all of the stars in the LMC have formed during the last 4 Gyr.

## 4. DISCUSSION

Taken at face value the results discussed above suggest that the rate of cluster formation in the Large Cloud increased 3-5 Gyr ago by a much larger factor than did the rate of star formation. Van den Bergh (1979) tentatively suggested that this difference between the present rates of cluster formation (normalized to the same rates of star formation) in the active LMC, and in the more quiescent dwarf IC 1613, might be due to a difference in the frequency with which strong shocks occur in these two galaxies. This idea was originally motivated by the observation that the nucleus of the active galaxy M 82 is embedded in a swarm of super-luminous clusters. Tinsley (1979) had also suggested that strong shocks might promote the formation of clusters. A possible argument against this hypothesis is that the Fornax dwarf spheroidal galaxy has a high specific globular cluster frequency, even though its internal velocity dispersion is low. However, support for the idea that strong shocks promote cluster formation has recently been provided by the observation that gas-rich merging galaxies, such as NGC 4038/39 "the antennae" (Whitmore & Schweizer 1995), and the recent merger remnant NGC 7252 (Miller et al. 1997) contain large numbers of luminous young star clusters. If this hypothesis is correct then the high rate of globular cluster formation in the Large Cloud ~ 13 Gyr ago might also be due violent cloud-cloud collisions, resulting in strong shocks, during the earliest evolutionary phase of the LMC. The Large Cloud contains 13 old globular clusters and is 5.8 times more luminous than the Small Cloud. One might, other things being equal, therefore have expected ~ 2.2 old globulars in the SMC. In fact, none are observed. NGC 121, which is the only true globular cluster in the SMC, is significantly younger than the old LMC clusters. The absence of such old globulars in the SMC would be consistent with the suggestion that this object formed more gradually by gentle subsidence and merger of ancestral fragments. However, it might also be argued that this observed difference resulted from the well-known perversity of small-number statistics. Harris (1991) finds that the specific globular cluster frequency in the outer part of the giant elliptical galaxy M 49 is an order of magnitude higher than it is in the core of this galaxy. Taken at face value this suggests that the rate of cluster formation, normalized to the same rate of star formation, was greater in the outer halo of M 49 than it was in the core of this galaxy.

In summary it is concluded that Baade was probably too optimistic when he suggested that the history of star formation in galaxies can be derived from their history of cluster formation. In fact it appears that the rate of cluster formation, normalized to the rate of star formation, can differ significantly from galaxy to galaxy. Within an individual galaxy it can depend on position and can change with time. It is speculated that strong shocks favor the formation of clusters. The reason for the sudden increase in the normalized frequency of cluster formation in the LMC 3-5 Gyr ago remains unknown. Possibly the relatively recent formation of the Large Cloud Bar generated strong shocks in the LMC gas. These shocks might then have triggered the formation of numerous clusters.

I wish to express my thanks to a particularly helpful anonymous referee.

# REFERENCES

Baade, W. 1963, Evolution of Stars and Galaxies (Cambridge: Harvard University Press)

Butcher, H. 1977, ApJ, 216, 372

Da Costa, G.S. 1991, in The Magellanic Clouds = IAU Symposium No. 148, Eds. R. Haynes and D. Milne (Dordrecht: Kluwer), p. 183

Fukui, Y. et al. 1999, in New Views of the Magellanic Clouds = IAU Symposium No. 190, Eds. Y.-H. Chu, J.E. Hesser & N.B. Suntzeff (San Francisco: ASP), in preparation

Geisler, D., Bica, E., Dottori, H., Claría, J.J., Piatti, A.E. & Santos, J.F.C. 1997, AJ, 114, 1920

Harris, W.E.1 991, ARA&A, 29, 543

Hodge, P.W. 1988, PASP, 100, 576

Hodge, P.W. 1998, Colloquium given in Victoria, B.C. 1998, April 21

Johnson, J.A., Bolte, M., Bond, H.E., Hesser, J.E., de Oliveira, C.M., Richer, H.B., Stetson, P.B. & VandenBerg, D.A. 1999, in New Views of the Magellanic Clouds = IAU Symposium No.190, Eds. Y.-H. Chu, J.E. Hesser & N.B. Suntzeff (San Francisco: ASP), in preparation

Mighell, K.J., Sarajedini, A. & French, R.S. 1998, AJ, in preparation

Mighell, K.J., Sarajedini, A. & French, R.S. 1999, in New Views of the Magellanic Clouds = IAU Symposium No. 190, Eds. Y.-H. Chu, J.E. Hesser & N.B. Suntzeff (San Francisco: ASP), in preparation

Miller, B.W., Whitmore, B.C., Schweizer, F. & Fall, S.M. 1997, AJ, 114, 2381

Olsen, K.A.G., Hodge, P.W., Mateo, M., Olszewski, E.W., Schommer, R.A., Suntzeff, N.B. & Walker, A.R. 1999a, in New Views of the Magellanic Clouds = IAU Symposium No. 190, Eds. Y.-H. Chu, J.E. Hesser & N.B. Suntzeff (San Francisco: ASP), in preparation

Olsen, K.A.G., Hodge, P.W., Mateo, M., Olszewski, E.W., Schommer, R.A., Suntzeff, N.B. & Walker, A.R. 1999b, MNRAS, in press

Sarajedini, A. 1998, AJ, 116, xxx

Sarajedini, A. 1999, in New Views of the Magellanic Clouds = IAU Symposium No. 190, Eds. Y.-H. Chu, J.E. Hesser & N.B. Suntzeff (San Francisco: ASP), in preparation

Stryker, L.L. 1984, ApJS, 55, 127

Tinsley, B.M. 1979, in Large-Scale Characteristics of the Galaxy = IAU Symposium No. 84, Ed. W.B. Burton (Dordrecht: Reidel), p. 431

van den Bergh, S. 1979, ApJ, 230, 95

van den Bergh, S. & McClure, R.D. 1980, A&A, 88, 360

Whitmore, B.C. & Schweizer, F. 1995, AJ, 109, 960